\documentclass[referee]{raa}            

\usepackage{graphicx,times}             
\usepackage{natbib}
\usepackage{amssymb,amsmath}
\bibpunct{(}{)}{;}{a}{}{,}

\usepackage[a4paper=true,dvipdfm=true,pagebackref=true]{hyperref}
\hypersetup{colorlinks = true, linkcolor = green, anchorcolor = red, citecolor = blue, filecolor = red, pagecolor = red, urlcolor = red}

\begin{document}

   \title{Magneto-Acoustic Waves in Magnetic Twisted Flux Tube}

   \volnopage{Vol.0 (20xx) No.0, 000--000}      
   \setcounter{page}{1}          

   \author{Wei Wu \inst{1,2}, Robert Sych\inst{2,3}, Jie Chen\inst{2} and *Jiang-tao Su\inst{2,1}}

   \institute{School of Astronomy and Space Sciences, University of Chinese Academy of Sciences 
19 A Yuquan Road, Shijingshan District, Beijing 100049, China\\
        \and
             Key Laboratory of Solar Activity, National Astronomical Observatories, Chinese
Academy of Sciences, Beijing 100101, China; {\it sjt@nao.cas.cn}\\
        \and 
             Institute of Solar-Terrestrial Physics SB RAS, Irkutsk 664033, Russia\\
\vs\no
   {\small Received~~20xx month day; accepted~~20xx~~month day}}

\abstract{At present, many works about MHD wave diagnostics in magnetic flux tubes are based on some pioneer works not considering the contributions of magnetic twist. Other works considered the effect on MHD waves, but the dispersion relationship they presented only gave the wave modes of $m=0,1,2...$. The kink mode of $m=-1$ was absent. Therefore, in this work we would like to present a complete dispersion relationship that includes both magnetic twist and the wave mode of $m=-1$. Analogous to the $m=+1$ wave mode, the mode of $m=-1$ also exhibits the mode change at finite $kr_{0}$, from body to surface mode. The phase speeds of this mode are usually less than those of $m=+1$ mode. The harmonic curves of $m=\pm1$ modes in dispersion relationship diagrams are approximately symmetric in respect to a characteristic velocity, e.g. the tube velocity in flux tubes. Based on the present dispersion relationship, we revist the issue of spiral wave patterns in sunspot and find that the magnetic twist has no great influence on their morphology in the frame of linear perturbation analysis.
}

   \authorrunning{W. Wu et al.}            
   \titlerunning{Magneto-Acoustic Waves}  

   \maketitle

%
%
\section{Introduction}           
\label{sect:intro}
In the context of MHD linear theory, wave propagation in untwisted magnetic flux tubes embeded in a magnetic environment was investigated by \cite{edw+1983}. The dispersion relations they derived have become a fundamental reference for the magneto-seismological inversion to probe the solar plasma. Later, \cite{ben+1999} extented their work and studied wave propagation in the twisted magnetic flux tubes in an incompressible medium. Due to the twist introduced, the hybrid (surface-body) modes of oscillation appear in contrast to the case of an untwisted incompressible tube. Sausage ($m = 0$) MHD wave propagation in incompressible (\citealt{erd+2006}) and compressible (\citealt{erd+2007}) magnetically twisted flux tubes was investigated further. More recent, \cite{erd+2010} studied the oscillatory modes of a magnetically twisted compressible flux tube in cylindrical geometry, and a general dispersion equation obtained in terms of Kummer functions for the approximation of weak and uniform internal twist.

However, it should be pointed out that \cite{ben+1999} and \cite{erd+2010} only presented the dispersion relationship diagrams (phase-speed diagrams of MHD oscillation modes) for the $m=+1$ mode of oscillation that will increase the twist in the tube. However, neither of them presented the $m=-1$ one of oscillation. This is because that the $m = 1$ mode is the most unstable of the two since the helical perturbation formed is of the same sense as the twist (\citealt{ben+1999}). 
With more advanced astronomical instruments put into service, more fine structures of MHD waves on the Sun are detected, e.g. the armed-spiral wavefronts in the umbrae of sunspots (e.g., \citealt{sych+2014}; \citealt{su+2016}; \citealt{sych+2020}). They are likely created by the superpsition of non-zero azimuthal modes driven 1600 km below the photosphere in the sunpots (\citealt{kang+2019}). For example, the one-armed pattern is produced by the slow-body sausage ($m=0$) and kink ($m=1$) modes. Naturally, we wander what the wave pattern looks like if magnetic twist is included in the flux tubes or the $m=-1$ mode is in place of that of $m=+1$. This is our motivation to complete the dispersion relationship diagram by including the $m=-1$ mode.

In this paper, we attempt to give a general dispersion equation for the $m=-1$ oscillation mode in Section 2. Complete diagrams of the dispersion relations are described in Section 3 and modeling of spiral wave patterns in Section 4. Finally, we conclude in Section 5.

\section{General dispersion equation of the allowed eigenwodes}

\label{sect:model}
\subsection{Magnetic twisted flux tube model}
We consider the modes of oscillation of a compressible magnetized twisted flux tube embedded within a uniformly magnetized plasma environment in cylindrical geometry ($r,\theta,z$) (\citealt{erd+2010}). In equilibrium, the plasma and magnetic field pressure satisfy the condition in the radial direction

\begin{equation}
\begin{array}{ll}
\frac{d}{dr}(p_{0}+\frac{B^2_{0}}{2\mu})=-\frac{B^2_{0\theta}}{\mu_{0}{r}},\\
\end{array}
\end{equation}
Here $B_{0} = (B^2_{0\theta} + B^2_{0z})^{\frac{1}{2}}$ denotes the strength of the equilibrium magnetic
field and $\mu_{0}$ is the magnetic permeability. The plasma density is taken to be uniform.

\label{sect:Det}
\subsection{Dispersion equation}
A general dispersion equation has been obtained in terms of Kummer functions for the magneto-acoustic waves in the compressible magnetized twisted flux tubes (\citealt{erd+2010}), which is

\begin{equation}
\begin{array}{ll}
D_\mathrm{e}\frac{r_{0}}{m_\mathrm{0e}}\frac{K_{m}(m_\mathrm{0e}r_{0})}{K'_{m}(m_\mathrm{0e}r_{0})}=-\frac{A^{2}r^{2}_{0}}{\mu^{2}_{0}}+D_\mathrm{i}r^{2}_{0}\frac{(1-\alpha^{2})}{m(1-\alpha)+2x_{0}\frac{M'(a,b,x_{0})}{M(a,b,x_{0})}},\\
\end{array}
\end{equation}
where the subscripts $\mathrm{i}$ and $\mathrm{e}$ respectively represent the interior and exterior of the tube, and $r_{0}$ is the radius of the tube, $m$ is azimuthal order of the mode, $b=m+1$, $A$ is arbitrary constant, $K_{m}$ is the modified Bessel functions of the second kind, $K'_{m}$ is its derivative,
$M(a,b,x_{0})$ is the Kummers function evaluated at $x = x_{0}$, $M'(a,b,x_{0})=\frac{a}{b}M(a+1,b+1,x_{0})$ is its derivative, and 

\begin{equation}
\begin{array}{ll}
m_\mathrm{0e}=\sqrt{\frac{(k^{2}C^{2}_\mathrm{Se}-\omega^{2})(k^{2}V^{2}_\mathrm{Ae}-\omega^{2})}{(V^{2}_\mathrm{Ae}+C^{2}_\mathrm{Se})(k^{2}C^{2}_\mathrm{Te}-\omega^{2})}},\\
\end{array}
\end{equation}
where $C_\mathrm{Se}$ and $V_\mathrm{Ae}$ are the external acoustic and Alfv$\acute{e}$n speeds, respectively, and 
$C_\mathrm{Te}=\sqrt{\frac{V^{2}_\mathrm{Ae}C^{2}_\mathrm{Se}}{V^{2}_\mathrm{Ae}+C^{2}_\mathrm{Se}}}$ is the external tube speed. For the meanings of other parameters, please refer to \cite{erd+2010}. 

In Equation (1), negative order $m$ is not allowed as $m=-1, -2, ...$ and $b=0, -1, ...$ are the forbidden values of the denominatorial parameters of Kummer function (see 47:9 in \citealt{old+2009}). In this case, it may be approximately proportional to another Kummer 
function. When $m=-1$, $b=0$ and $bM(a,b,x_{0})\approx{ax_{0}M(a+1,2,x_{0})}$. Equation (1) reduces to 

\begin{equation}
\begin{array}{ll}
D_\mathrm{e}\frac{r_{0}}{m_\mathrm{0e}}\frac{K_{1}(m_\mathrm{0e}r_{0})}{K'_{1}(m_\mathrm{0e}r_{0})}=-\frac{A^{2}r^{2}_{0}}{\mu^{2}_{0}}+D_\mathrm{i}r^{2}_{0}\frac{(1-\alpha^{2})}{(\alpha-1)+2\frac{M(a+1,1,x_{0})}{M(a+1,2,x_{0})}}.\\
\end{array}
\end{equation}
When $m=-2$, $b=-1$, $bM(a,b,x_{0})\approx{\frac{1}{2}M(a,1,x_{0})}$ and $M'(a,b,x_{0})=-aM(a+1,0,x_{0})\rightarrow\infty$. Equation (1) reduces to
 
\begin{equation}
\begin{array}{ll}
D_{e}\frac{r_{0}}{m_\mathrm{0e}}\frac{K_{1}(m_\mathrm{0e}r_{0})}{K'_{1}(m_\mathrm{0e}r_{0})}=-\frac{A^{2}r^{2}_{0}}{\mu^{2}_{0}}.\\
\end{array}
\end{equation}

On the other hand, \cite{ben+1999} obtained an exact dispersion relation for wave propagation in incompressible twisted magnetic flux tubes, in which the modified Bessel functions $I_{m}$ and $K_{m}$ accept a negative order $m$. Therefore, no more efforts should be done to change it.

\subsection{Stability of the compressible twisted flux tubes} 
We would like to check stability of the twisted flux tubes when they are disturbed slightly. In Equation (3), let $\omega^{2}=0$, then $m_\mathrm{0e}=k$. Now, $a\rightarrow\infty$ and $x_{0}=\frac{1}{4}\frac{r^{2}_{0}k^{2}_{\alpha}}{a}\rightarrow0$, where we have written

\begin{equation}
\begin{array}{ll}
k_{\alpha}=k\sqrt{1-\alpha^2},\\
\end{array}
\end{equation}

\begin{equation}
\begin{array}{ll}
\alpha^2=\frac{4A^2}{\mu_{0}\rho_\mathrm{0i}\omega^2_\mathrm{Ai}},\\
\end{array}
\end{equation}

\begin{equation}
\begin{array}{ll}
\omega_\mathrm{Ai}=\frac{A}{\sqrt{\mu\rho_\mathrm{0i}}}(m+kp),\\
\end{array}
\end{equation}

\begin{equation}
\begin{array}{ll}
p=\frac{B_{0z}}{A}.\\
\end{array}
\end{equation}
Note that we have dropped $2\pi$ in the last equation that represents the pitch of the magnetic field. Then, 

\begin{equation}
\begin{array}{ll}
\lim_{a\to\infty, x_{0}\to 0}2x_{0}\frac{M'(a,b,x_{0})}{M(a,b,x_{0})}=k_{\alpha}r_{0}\frac{I_{m+1}(k_{\alpha}r_{0})}{I_{m}(k_{\alpha}r_{0})},\\
\end{array}
\end{equation}
in Equation (1) and it becomes 

\begin{equation}
\begin{array}{ll}
D_\mathrm{e}\frac{r_{0}}{k}\frac{K_{m}(kr_{0})}{K'{m}(kr_{0})}=-\frac{A^{2}r^{2}_{0}}{\mu^{2}_{0}}+D_\mathrm{i}r^{2}_{0}\frac{(1-\alpha^{2})}{m(1-\alpha)+k_{\alpha}r_{0}\frac{I_{m+1}(k_{\alpha}r_{0})}{I_{m}(k_{\alpha}r_{0})}},\\
\end{array}
\end{equation}
where $D_\mathrm{e}=-\frac{1}{\mu_{0}}A^{2}(m+kp)^2$ and $D_\mathrm{i}=-\frac{1}{\mu_{0}}k^2B^2_\mathrm{e}$ as $\omega^{2}=0$. In form, the equation is consistent with the dispersion equation of the oscillation modes of a twisted magnetic flux tube in an incompressible medium (\citealt{erd+2010}). We further define 

\begin{equation}
\begin{array}{ll}
\mathcal{K}_{m}=\frac{kr_{0}K'_{m}(kr_{0})}{K_{m}(kr_{0})}, \chi_{m}=k_{\alpha}r_{0}\frac{I_{m+1}(k_{\alpha}r_{0})}{I_{m}(k_{\alpha}r_{0})}, \\
\end{array}
\end{equation}
and Equation (10) becomes

\begin{equation}
\begin{array}{ll}
[(m+kp)\chi_{m}+2m][1-(kr_{0})^2\frac{(B_\mathrm{e}/Ar_{0})^2}{\mathcal{K}_{m}}]=4(m+kp)-(m+kp)^3, \\
\end{array}
\end{equation}
which is just Equation (35) in \cite{ben+1999}. Therefore, their analyses about stability of the disturbed incompressible twisted flux tube is also suitable for the case of compressibility. They argued that $m=\pm1$ are the most important modes and $m=1$ is the most unstable of the two. In this case, for small $kr_{0}$ and $kp$, when the ratio of the external field to azimuth field is unless than $\frac{1}{2}$, that is $\frac{B_\mathrm{e}}{Ar_{0}} \ge \frac{1}{2}$, the tube is stable. In a following section, we only present the diagrams of dispersion relation for the $m=\pm1$ modes.

\section{Diagrams of the dispersion relations}
\label{sect:Diagram}
\subsection{Compressibility}
\begin{figure}
   \centerline{\includegraphics[width=15cm,clip=]{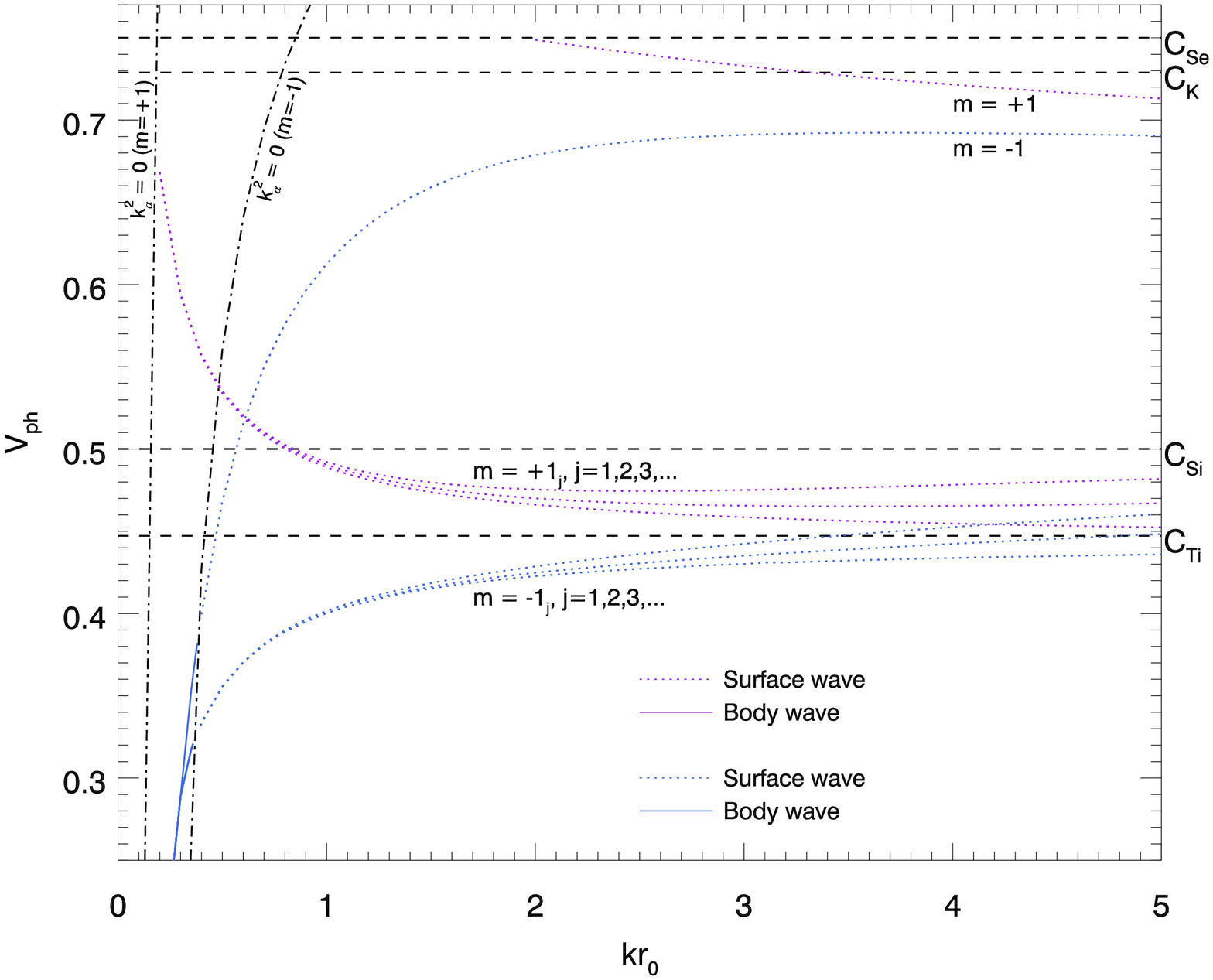}}
\caption{Under photospheric conditions ($C_\mathrm{Se} = 0.75V_\mathrm{Ai}$, $V_\mathrm{Ae} = 0.25V_\mathrm{Ai}$ and $C_\mathrm{Si} = 0.5V_\mathrm{Ai}$, please see \cite{erd+2010}), the diagram curves of the dimensionless phase speed ($V_\mathrm{ph}$) of the kink ($m=\pm1$) modes as function of the dimensionless wavenumber ($kr_{0}$) for a uniformly twisted intense magnetic flux tube ($V_\mathrm{Ai\phi} = 0.1$). $C_\mathrm{K}$, $C_\mathrm{Si}$ ($C_\mathrm{Se}$) and $C_\mathrm{Ti}$ are the kink, sound and tube characteristic speeds on the photosphere. The dot-dashed curves correspond to the place where $k^{2}_{\alpha}=0 (m=\pm1)$ in the plots. Their left-/right-side regions are the domains where only body ($k^{2}_{\alpha} < 0$)/surface ($k^{2}_{\alpha} > 0$) waves are present.}
\label{fig1}
\end{figure}

So far, the dispersion relation diagrams for the sausage ($m = 0$), kink ($m = 1$) and fluting ($m > 1$) modes have been extentively investigated, but those of their counterparts, the negative modes, are absent. We attempt to accomplish them and focus on the kink $m=\pm1$ wave modes. In a compressible medium for the $m=+1$ mode, \cite{erd+2010} demonstrated that under photsphere conditions, the fast kink surface mode has a cut-off at a phase speed $V_\mathrm{ph} = C_\mathrm{Se}$ that has no phase speed solution for small dimensionless wavenumbers, i.e. $kr_{0} << 1$ when there is magnetic twist, and the phase speed $V_\mathrm{ph}$ of the slow kink suface mode tends to infinity as $kr_{0} \to 0$ as shown in Figure \ref{fig1}. For the $m=-1$ mode, the figure shows that its phase speeds of both the slow and fast kink surface modes are less than those of the $m=+1$ mode, and of them the curves of the slow kink surface mode seem to be a mirror of the counterparts of the $m=+1$ mode relative to a speed slightly greater than $C_\mathrm{Ti}$. Moreover, there is a so-called mixed (hybrid) character at $kr_{0}\approx0.35$ for both the fast and slow surface kink modes, and their phase speed $V_\mathrm{ph}$ tends to zero as $kr_{0} \to 0$. With $kr_{0} \to 0$ the phase speed difference between the two modes of $m=\pm1$ becomes more and more significant. However, this difference would become less and less significant with $kr_{0}$ increasing. In Figure 2, we plot the normalized eigenfunctions $\xi_{r}$ and $P_\mathrm{T}$ of the fundamental $m = \pm1$ modes as function of radius $r_{0}$. Comparing the left two panels, we can find that the phase speed difference between the two modes is larger and their difference in amplitude ($\xi_{r}$) is also larger. However, their difference in total pressure disturbance ($P_\mathrm{T}$) does not show such similar change as shown in the right two panels.

These features will also appear in the following diagrams of dispersion relation. Under coronal conditions, as shown in Figure \ref{fig3} the phase speed $V_\mathrm{ph}$ of the fask kink surface mode of $m=-1_{1}$ is a little greater than that of the fast kink surface mode of $m=+1_{1}$, the two phase speeds of the fast kink surface mode of $m=\pm1_{2}$ are equal in magnitude, and the phase speed $V_\mathrm{ph}$ of the fask kink surface mode of $m=-1_{3}$ is less than that of the fast kink surface mode of $m=+1_{3}$. The phase speeds of the fast kink surface modes of $m=\pm1_{2, 3}$ have a cut-off at a phase speed $V_\mathrm{ph} = V_\mathrm{Ae}$. For the slow kink surface modes, both the phase speed curves of the $m=\pm1$ modes shows a mixed feature as $kr_{0} \to 0$, of which the $m=+1$ mode tends to infinity while the $m=-1$ mode tends to zero. Moreover, the two sets of curves show symmetry in morphology relative to a speed slightly greater than $C_\mathrm{Ti}$. In Figure 4, we plot the normalized eigenfunctions $\xi_{r}$ and $P_\mathrm{T}$ of the fundamental $m = \pm1$ modes as function of radius $r_{0}$. It shows that the phase speed difference being large between the two modes does not mean their differences in amplitude of $\xi_{r}$ and $P_\mathrm{T}$ is also large.

\subsection{Incompressibility}
In an incompressible medium, the sound speed tends to infinite and the slow-wave tube speed becomes the Alfv$\acute{e}$n speed. The fast waves are eliminated from the system and there is no distinction between the Alfv$\acute{e}$n continuum and the slow continuum (\citealt{ben+1999}). Following \cite{ben+1999}, we take external field $B_\mathrm{e}=0.5\sqrt{B^{2}_{0z}+B^{2}_{\theta}}$ and $B_{\theta}=0.1B_{0z}$. The dimensionless phase speed ($V_\mathrm{ph}$) of the kink modes ($m=\pm1$) as function of the dimensionless wavenumber ($kr_{0}$) for an incompressible twisted magnetic flux tube are shown in Figure \ref{fig5}. The phase velocity bands of the body waves of $m=\pm1$ show symmetry with respect to the Alfv$\acute{e}$n speed $V_\mathrm{Az}$. The $m=+1$ mode tends to infinity, while the $m=-1$ mode tends to $0$ as $kr_{0} \to 0$. The surface wave of $m=+1$ shows a mixed feature as $kr_{0} \to 0$ and that of $m=-1$ shows the feature at $kr_{0}\approx1.1$. In Figure 6, we plot the normalized eigenfunctions $\xi_{r}$ and $P_\mathrm{T}$ of the fundamental $m = \pm1$ modes as function of radius $r_{0}$. It shows that the differences of $\xi_{r}$ and $P_\mathrm{T}$ between the modes decrease with their phase speed differences decreasing. 

\section{Modeling of spiral wave patterns}
\label{sect:Modeling}
\cite{kang+2019} interpreted the spiral wave patterns (SWPs, \citealt{sych+2014}; \citealt{su+2016}; \citealt{sych+2020}) as the azimuthal wave modes ($m=0,1,2$) propagating in an untwisted uniform magnetic cylinder. In this work, we follow their senario and revisit the issue. The magnetic twist is added into the flux tube and the wave modes with $m=\pm1$ are investigated. The internally oscillatory solution (surface waves) of the longitudinal velocity is given as follows (\citealt{erd+2010})  
\begin{equation}
\begin{array}{ll}
v_{z}=-C_{1}{\{kr_{0}-\frac{V_{\mathrm{Ai}z}V^2_\mathrm{ph}(mV_\mathrm{Ai\phi}+kr_{0}V_{\mathrm{Ai}z})}{V^2_\mathrm{ph}(C^2_\mathrm{Si}+V^2_\mathrm{Ai})-C^2_\mathrm{Si}V^2_{f_{B}}}{[1-2V^2_\mathrm{Ai\phi}\frac{m(1-\alpha)+\frac{2xM'(a,b,x)}{M(a,b,x)}}{k^2r^2_{0}(V^2_\mathrm{ph}-V^2_{f_{B}})(1-\alpha^2)}]}\}}\frac{x^{\frac{m}{2}}exp^{-\frac{x}{2}}M(a,b,x)exp^{i(kz+m\phi-\omega t)}}{kr_{0}(V^2_\mathrm{ph}-V^{2}_{f_{B}})}, \\
\end{array}
\end{equation}
where $C_{1}$ is arbitrary constant, $V_\mathrm{ph}=\omega/k$ is phase speed, $V_{\mathrm{Ai}z}$ and $V_\mathrm{Ai\phi}$ are the two components of the internal Alfv$\acute{e}$n speed $V_\mathrm{Ai}$ in the longtudinal and azimuth directions of cylinder, respectively, $V_\mathrm{Si}$ is internal acoustic speed and 

\begin{equation}
\begin{array}{ll}
V_{f_{B}}=\sqrt{\frac{m^2V^2_\mathrm{Ai\phi}}{k^2r^2_{0}}+\frac{2mV_\mathrm{Ai\phi}V_{\mathrm{Ai}z}}{kr_{0}}+V^2_{\mathrm{Ai}z}}. \\
\end{array}
\end{equation}

The phase speed $V_\mathrm{ph}$ is derived from the dispersion relations (1) and (3) for the wave modes of $m=\pm1$, respectively. Under photosphere conditions, we take $V_\mathrm{Ai}=8182$ ms$^{-1}$, $C_\mathrm{Se} = 1.2V_\mathrm{Ai}$, $V_\mathrm{Ae} = 0.25V_\mathrm{Ai}$, $C_\mathrm{Si} = 1.1V_\mathrm{Ai}$, and $V_\mathrm{Ai\phi} = 0.1V_\mathrm{Ai}$. Table 1 lists all the relevant parameters related to the wave modes of $m=\pm1$ propagating in the twisted magnetic cylinder. For comparison, the $m=+1$ wave mode in untwisted cylinder is also included in the table. As $k^2_{\alpha}>0$ for the listed three modes, they are all surface waves which decay away from the cylinder's suface. 

Figure 7 shows the normalized $v_{z}$ as function of the radius of cylinder for the $m=\pm1$ wave modes in a twisted flux tube and $m=+1$ in an untwisted flux tube. At a fixed radius value, the value of $m=+1$ wave mode in the untwisted flux tube is less/greater than that of $m=-1$/$m=+1$ wave mode in the twisted flux tube. Figure 8 shows $v_{z}$ maps of the $m=\pm1$ wave modes emerging from a depth of $d=100$ km under photosphere. The $m=\pm1$ wave modes have an opposite phase relation. In morphology, there is no significant difference between the wave modes of $m=+1$ in the twisted and untwisted magnetic cylinders. It indicates that the magnetic twist has no great influence on the morphology of SWPs in the frame of linear perturbation analysis. We can suppose that the complex structure of magnetic field in sunspot with different cutoff frequencies along different polar angles play major role in forming SWPs. Waves propagate radially along magnetic waveguides will leads to their frequency fragmentation and the appearance of a quasi-spiral spatial shape. 

\section{Conclusions}
\label{sect:conclusion}
In this work, based on the work of \cite{erd+2010} we derive the dispersion equation for the $m=-1$ magneto-acoustic wave mode in the compressible magnetized twisted flux tubes. We find that the analyses about stability of the incompressible twisted flux tube are also suitable for the case of compressibility. $m=\pm1$ are the most important modes and $m=1$ is the most unstable of the two. We then present a complete dispersion relationship diagram that includes both the magnetic twist and the negative order $m=-1$. Furthermore, we revisit the issue of spiral wave patterns in sunspot and find that there is no obvious difference in the morphologies of umbral spiral waves with and without magnetic twist. The main results for the magneto-acoustic waves in magnetic twisted flux tube are summarized as follows.

(1) Under photosphere conditions, the azimuthal $m=\pm1$ wave modes show great difference in phase-speed at finite $kr_{0}$ for both the fast and slow mode waves. This difference tends to decrease with $kr_{0}$ increasing.

(2) Under coronal conditions, the azimuthal $m=\pm1$ wave modes show great difference in phase-speed at finite $kr_{0}$ only for the slow mode waves, while for the fast mode waves they show a finite difference no more than $10\%$ at the allowed values of $kr_{0}$ for the first 
harmonic curves ($m=\pm1_{1}$).

(3) The magnetic twist may has no great influence on the morphology of SWPs in the frame of linear perturbation analysis. We think that the spatial and frequency fragmentation of wavefronts as the combination of narrowband spherical and linear parts of the wavefronts can provide the observed spirality. Study of relationship between wave shapes and maps of the magnetic field inclination angles presented in \cite{sych+2020} confirm this assumption.

\begin{acknowledgements}
We are grateful to an anonymous referee for constructive suggestions. This work is supported by the Strategic Priority Research Program on Space Science, the Chinese Academy of Sciences (Grant No. XDA15320302, XDA15052200, XDA15320102), National Natural Science Foundation of China (Grant No. 11773038, U1731241, 11427803), and the 13th Five-year Informatization Plan of Chinese Academy of Sciences (Grand No. XXH13505-04). RS research was performed within the basic funding from FR program II.16, RAS program KP19-270, and supported by the Chinese Academy of Sciences President’s International Fellowship Initiative, Grant No. 2020VMA0032.
\end{acknowledgements}

\begin{table}
\bc
\begin{minipage}[]{100mm}
\caption[]{Parameters of 9 objects observed by YFOSC\label{tab1}}\end{minipage}
\setlength{\tabcolsep}{1pt}
\small
 \begin{tabular}{cccccccccccccccc}
  \hline\noalign{\smallskip}
Twist& $m$& $C_\mathrm{Si}$ & $C_\mathrm{Ai}$ &$C_\mathrm{Ai\phi}$&$C_\mathrm{Ae}$&$C_\mathrm{Se}$& $T_\mathrm{ph}$&$\omega$& $kr_{0}$& $V_\mathrm{ph}$&$k^2_{\alpha}$& $k$& $r_{0}$ & $d$\\
& &($ms^{-1}$)&($m s^{-1}$)&($m s^{-1}$)&($m s^{-1}$)&($m s^{-1}$)&($s$)&($rad s^{-1}$)&&($m s^{-1}$)&&($10^{-6}$)&(km)&(km)&\\
  \hline\noalign{\smallskip}
No &+1&9000&8182&818&2045&9818&150&0.042&3&5574&$>0$&7.51&399&100\\
Yes&+1&9000&8182&818&2045&9818&150&0.042&3&5724&$>0$&7.32&410&100\\
Yes&-1&9000&8182&818&2045&9818&150&0.042&3&5365&$>0$&7.81&384&100\\
  \noalign{\smallskip}\hline
\end{tabular}
\ec
\tablecomments{0.86\textwidth}{$T_\mathrm{ph}$ is period of the wave modes and $d$ is depth of the oscillation source under photosphere.}
\end{table}
;------------------------------------------------------------------------
\begin{figure}
   \centerline{\includegraphics[width=15cm,clip=]{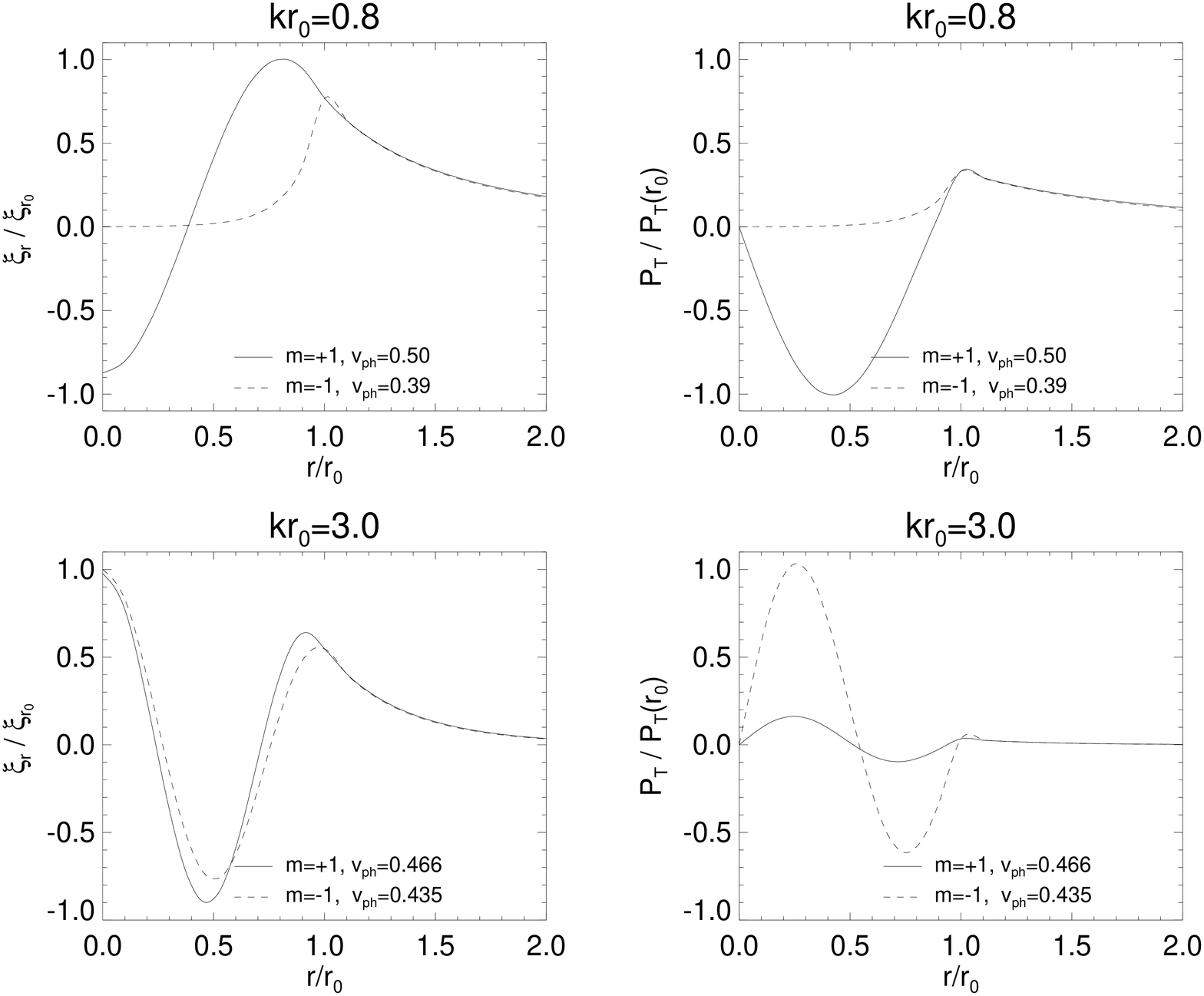}}
\caption{The normalised eigenfunctions $\xi_{r}/\xi_{r_{0}}$ (left panels) and $P_\mathrm{T}/P_\mathrm{T}(r_{0})$ (right panels) of the kink $m =\pm1$ modes 
are plotted for the case of $C_\mathrm{Se} = 0.75V_\mathrm{Ai}$, $V_\mathrm{Ae} = 0.25V_\mathrm{Ai}$ and $C_\mathrm{Si} = 0.5V_\mathrm{Ai}$ (photosphere conditions) when $V_\mathrm{Ai\phi} = 0.1$ in a compressible
medium.}
\label{fig2}
\end{figure}
\begin{figure}
\centerline{\includegraphics[width=15cm,clip=]{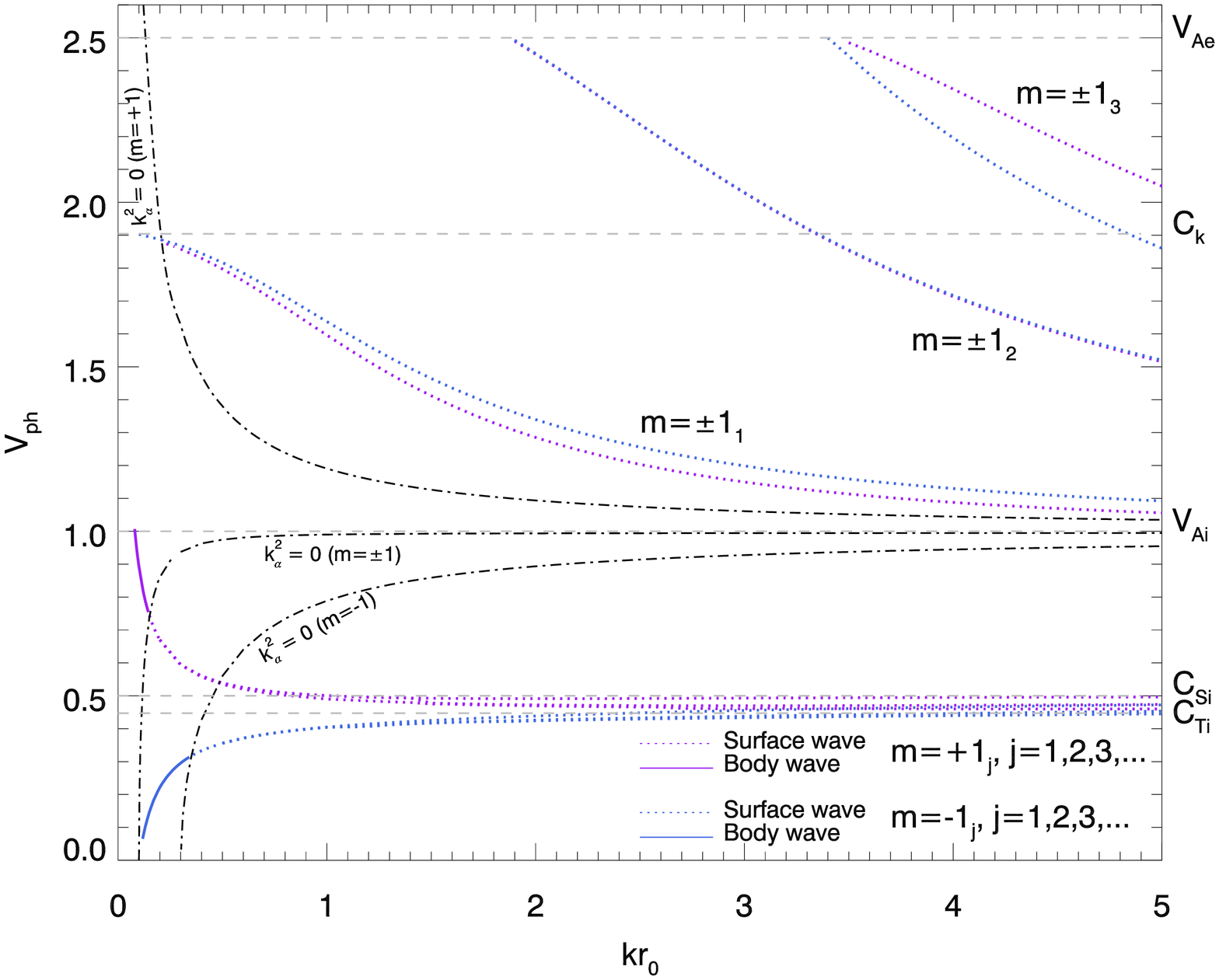}}
\caption{Under coronal conditions ($C_\mathrm{Se} = 0.25V_\mathrm{Ai}$, $V_\mathrm{Ae} = 2.5V_\mathrm{Ai}$ and $C_\mathrm{Si} = 0.5V_\mathrm{Ai}$, please see \cite{erd+2010}), the diagram curves of the dimensionless phase speed ($V_\mathrm{ph}$) of the kink ($m=\pm1$) modes as function of the dimensionless wavenumber ($kr_{0}$) for a uniformly twisted intense magnetic flux tube ($V_\mathrm{Ai\phi} = 0.1$). $V_\mathrm{Ai}$ ($V_\mathrm{Ae}$), $C_\mathrm{K}$, $C_\mathrm{Si}$ ($C_\mathrm{Se}$) and $C_\mathrm{Ti}$ are the Alfv$\acute{e}$n, kink, sound and tube characteristic speeds on the corona. The dot-dashed curves correspond to the place where $k^{2}_{\alpha}=0$ in the plots. Note that the top one is for the $m=+1$ mode, the bottom for $m=-1$ mode and the middle for the both. For each mode ($m=+1$ or $m=-1$), the body ($k^{2}_{\alpha} < 0$) waves only exist in between curves.}
\label{fig3}
\end{figure}
\begin{figure}
\centerline{\includegraphics[width=15cm,clip=]{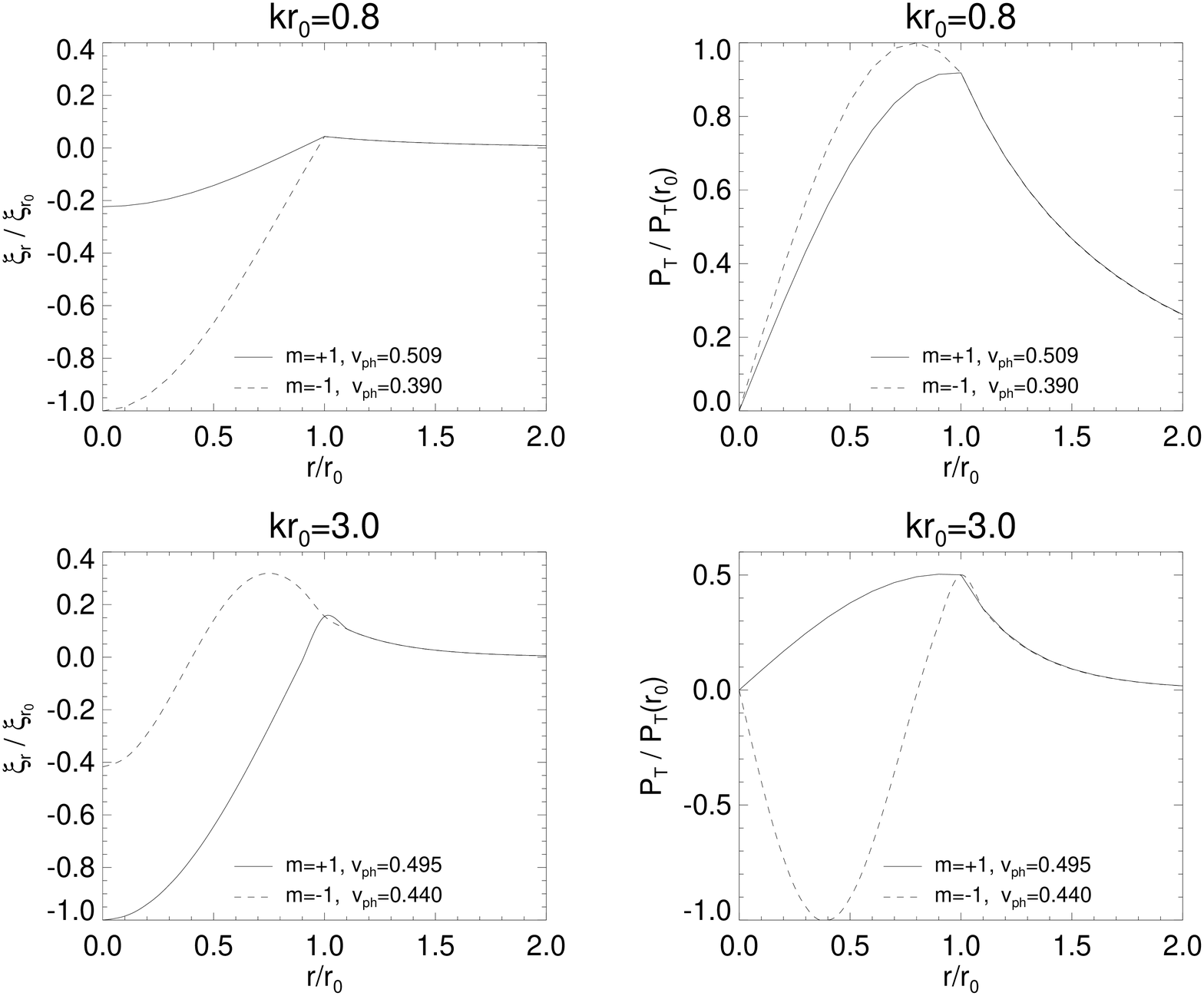}}
\caption{The normalised eigenfunctions $\xi_{r}/\xi_{r_{0}}$ (left panels) and $P_\mathrm{T}/P_\mathrm{T}(r_{0})$ (right panels) of the kink $m =\pm1$ modes 
are plotted for the case of $C_\mathrm{Se} = 0.25V_\mathrm{Ai}$, $V_\mathrm{Ae} = 2.5V_\mathrm{Ai}$ and $C_\mathrm{Si} = 0.5V_\mathrm{Ai}$ (coronal conditions) when $V_{Ai\phi} = 0.1$ in a compressible medium.}
\label{fig4}
\end{figure}
\begin{figure}
\centerline{\includegraphics[width=15cm,clip=]{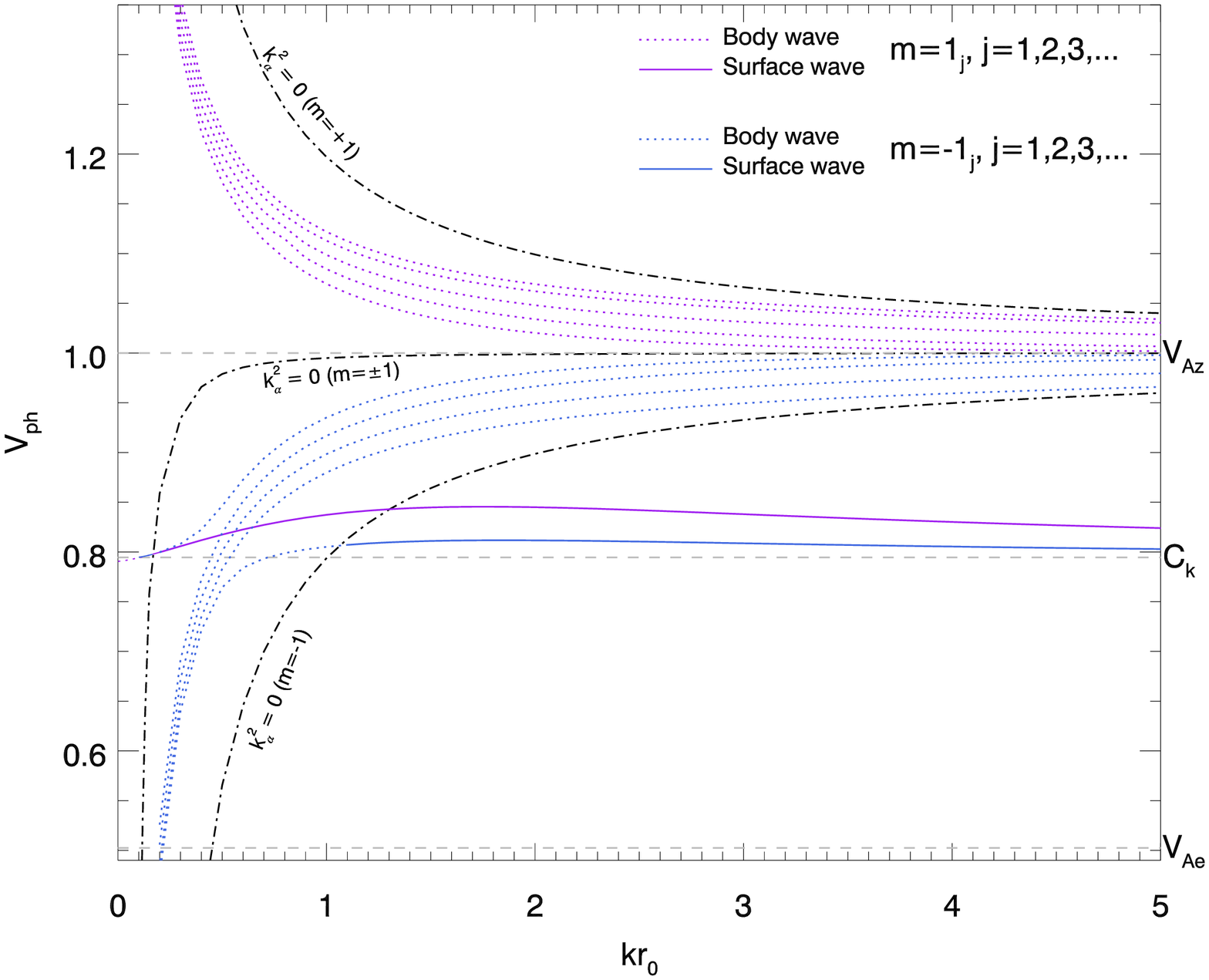}}
\caption{The diagram curves of the dimensionless phase speed ($V_\mathrm{ph}$) of the kink modes ($m=\pm1$) as function of the dimensionless wavenumber ($kr_{0}$) for an uncompressible twisted magnetic flux tube when $B_{\theta}=0.1B_{0z}$ and $B_\mathrm{e}=0.5\sqrt{B^{2}_{0z}+B^{2}_{\theta}}$ (\citealt{ben+1999}). $C_\mathrm{K}$ and $V_\mathrm{Ai}/V_\mathrm{Ae}$ are the kink and Alfv$\acute{e}$n characteristic speeds. The dot-dashed curves correspond to the place where $k^{2}_{\alpha}=0$ in the plots. Note that the top one is for the $m=+1$ mode, the bottom for $m=-1$ mode and the middle for the both. For each mode ($m=+1$ or $m=-1$), the body ($k^{2}_{\alpha} < 0$) waves only exist in between curves.}
\label{fig5}
\end{figure}
\begin{figure}
\centerline{\includegraphics[width=15cm,clip=]{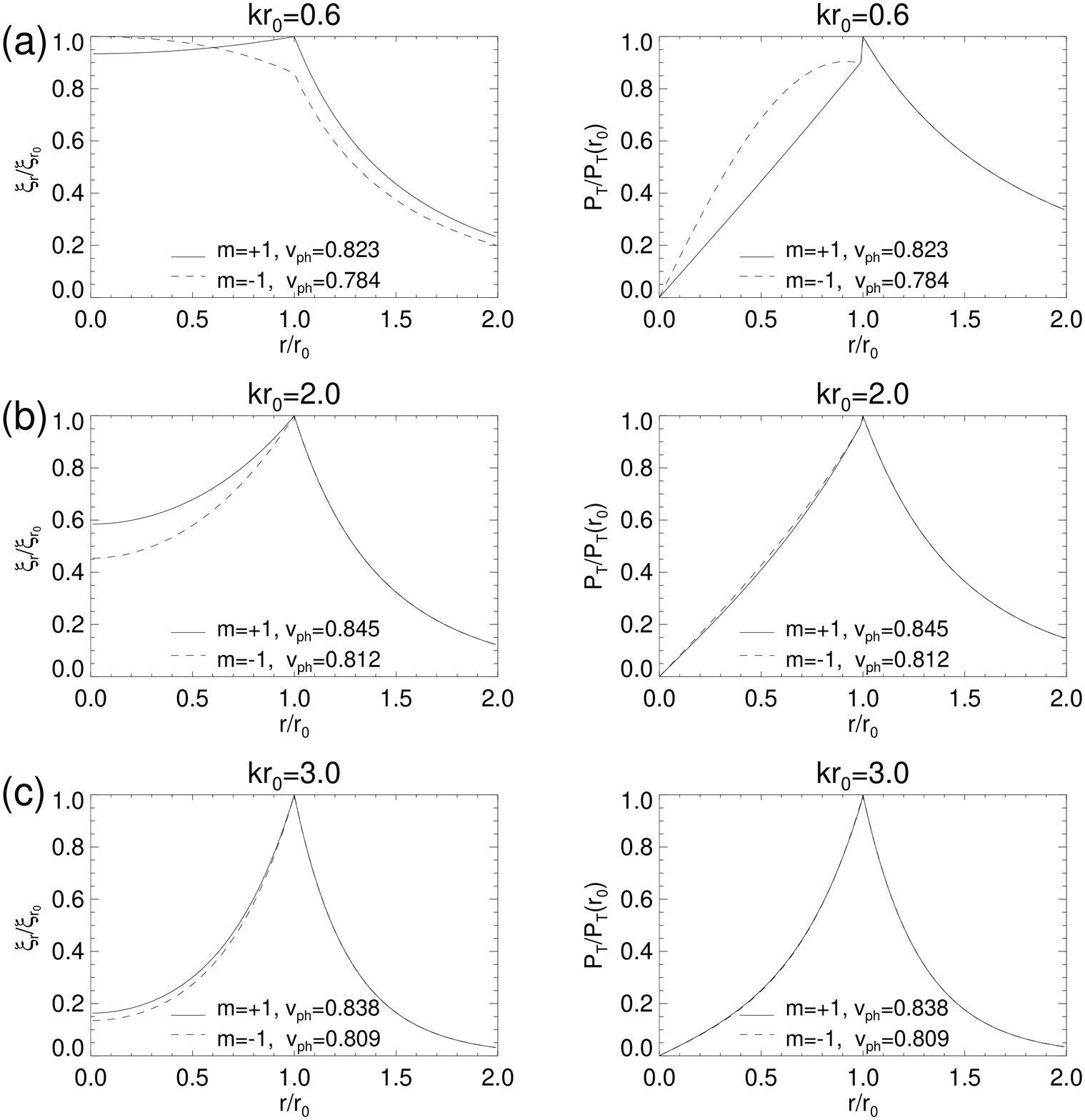}}
\caption{The normalised eigenfunctions $\xi_{r}/\xi_{r_{0}}$ (left panels) and $P_\mathrm{T}/P_\mathrm{T}(r_{0})$ (right panels) of the kink $m =\pm1$ modes 
are plotted for the case $B_{\theta}=0.1B_{0z}$ and $B_\mathrm{e}=0.5\sqrt{B^{2}_{0z}+B^{2}_{\theta}}$ in an incompressible
medium.}
\label{fig6}
\end{figure}
\begin{figure}
\centerline{\includegraphics[width=10cm,clip=]{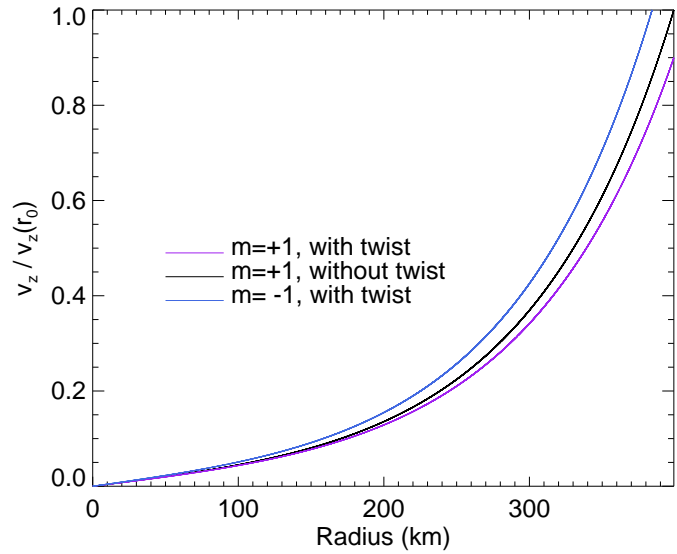}}
\caption{The normalised velocity componnet $v_{z}/v_{z}(r_{0})$ of the kink $m =\pm1$ wave modes as function of radius are plotted 
in an incompressible magnetic sylinder, of which the blue/pink curves are for the twisted sylinder and the black curve for the untwisted one.}
\label{fig7}
\end{figure}
\begin{figure}
\centerline{\includegraphics[width=15cm,clip=]{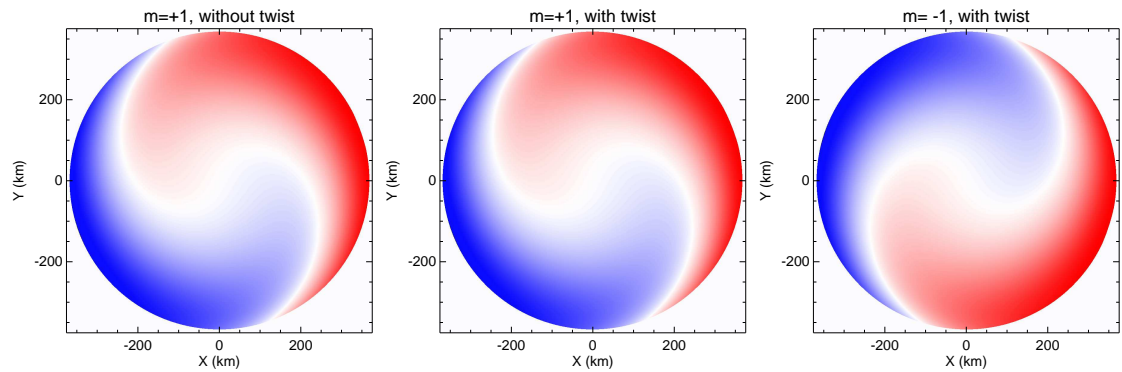}}
\caption{Snapshots of the simulated velocity compnent of $v_{z}$ for the kink $m=\pm1$ wave modes in ${x-y}$ plane. The left panel is for the untwisted magnetic sylinder and the middle and right ones for the twisted sylinder. An animation of this figure is available.}
\label{fig8}
\end{figure}

\label{lastpage}


\begin{thebibliography}{99}

  \bibitem[Bennett et al.(1999)]{ben+1999} Bennett, K., Roberts, B., \& Narain, U.\ 1999, \solphys, 185, 41
  
  \bibitem[Edwin \& Roberts (1983)]{edw+1983} Edwin, P.~M. \& Roberts, B.\ 1983, \solphys, 88, 179
   
  \bibitem[Erd{\'e}lyi \& Fedun(2006)]{erd+2006} Erd{\'e}lyi, R. \& Fedun, V.\ 2006, \solphys, 238, 41

  \bibitem[Erd{\'e}lyi \& Fedun(2007)]{erd+2007} Erd{\'e}lyi, R. \& Fedun, V.\ 2007, \solphys, 246, 101
  
  \bibitem[Erd{\'e}lyi \& Fedun(2010)]{erd+2010} Erd{\'e}lyi, R. \& Fedun, V.\ 2010, \solphys, 263, 63

  \bibitem[Kang et al.(2019)]{kang+2019} Kang, J., Chae, J., Nakariakov, V.~M., et al.\ 2019, \apjl, 877, L9

  \bibitem[Oldham et al. (2009)]{old+2009} Oldham, K., Myland, J. \& Spanier, J. \ 2009, An atlas of functions, 2nd edn. Springer, New York

  \bibitem[Su et al.(2016)]{su+2016} Su, J.~T., Ji, K.~F., Cao, W., et al.\ 2016, \apj, 817, 117

  \bibitem[Sych \& Nakariakov(2014)]{sych+2014} Sych, R. \& Nakariakov, V.~M.\ 2014, \aap, 569, A72

  \bibitem[Sych et al.(2020)]{sych+2020} Sych, R., Jess, D.~B., \& Su, J.\ 2020, Phil. Trans. R. Soc. A 379: 20200180
 
\end{thebibliography}
\end{document}